\documentstyle[aps,preprint]{revtex}
\begin {document}
\draft

\title{Precise solution of few-body problems with the stochastic variational
 method}

\author{K. Varga$^{1,2}$ and Y. Suzuki$^{3}$
\
\
\\ $^{1}$ RIKEN, Hirosawa, Wako, Saitama 351-01, Japan
\\ $^{2}$Institute of Nuclear Research of the Hungarian Academy of Sciences,
\\
Debrecen, P. O. Box 51, H--4001, Hungary
\\and
\\
$^{3}$Department of Physics, Niigata University, Niigata 950-21, Japan}
\date{\today}
\maketitle
\begin{abstract}
A precise variational solution to $N$=2--6-body problems is reported. The
trial wave functions are chosen to be combinations of
correlated Gaussians, which facilitate
a fully analytical calculation of the matrix elements. The nonlinear
parameters of the trial function are selected by a stochastic
method. Fermionic and bosonic few-body systems  are investigated for
interactions of different type.
A comparison of the results with those available in the literature shows
that the method is both accurate and efficient.
\end{abstract}
\vskip 1.cm
\narrowtext
\par\indent
The investigation of few-nucleon systems interacting via realistic forces
has always attracted much interest in nuclear physics.  Solving the
nuclear few-body Schr$\ddot{\rm o}$dinger equation is, however, extremely
difficult because of the pathological complexity of the interaction and of the
variety of nuclear
motion.  There has been considerable effort made to obtain accurate
ground-state
properties of the few-nucleon systems with variational \cite{crcbgv,atms},
Faddeev-Yakubovsky (FY) \cite{gloeckle,faddeev} and Quantum Monte Carlo
methods
\cite{mcv,gfm}. The Quantum Monte Carlo methods proved to be most successful by
being able to go beyond the four-nucleon problem.
The aim of this Letter is to present an alternative solution, a  variational
approach based on the stochastic variational method (SVM) \cite{SVM,VSL},
and to show its application to $N$=2--6-particle systems.
\par\indent
The variational foundation for the time-independent Schr\"odinger equation
provides a solid and arbitrarily improvable framework for the
approximate solution of bound-state problems. The most crucial point of the
variational approach is the choice of the trial function. There are
two widely applied strategies: (1) one selects the most appropriate functions
to describe the short-range as well as long-range correlations and calculates
the matrix elements by Monte Carlo technique, or (2) one uses simpler,
but possibly more, functions to facilitate the analytical calculation of the
matrix elements. We follow the second choice by using the correlated
Gaussian functions \cite{corrgauss} as spatial parts.
These functions for the $N$-nucleon system are given by
\begin{equation}
\Phi_{(LS)JMTT_z}({\bf x}, {\underline a})={\cal A} \lbrace
{\rm e}^{-{1 \over 2}{\bf x} {\underline a} {\bf x}} \left[ \theta_{L}({\bf x})
 \phi_S \right]_{JM}
\phi_{TT_z}\rbrace,
\end{equation}
where ${\bf x}= \{{\bf x}_1,...,{\bf x}_{N-1}\}$
is a set of Jacobi coordinates, $\underline a$ a positive-definite,
symmetric matrix of
 nonlinear
parameters, $\theta_{LM_L}({\bf x})$ a vector-coupled product of solid
 spherical
harmonics, and $\phi_{SM_S}$ and $\phi_{TT_z}$ are the spin-isospin parts of
the wave function. To simplify the notation, the quantum numbers of the
intermediate couplings are suppressed. The proper symmetry of the wave
function is effected by the antisymmetrizing (or, for bosons, symmetrizing)
operator ${\cal A}$. The trial function is then expressed as an expansion
over a number of functions of eq. (1).
 If these functions take into account
all important correlations within the system, then they can be said to form
a ``basis".
\par\indent
The variational approximation, however, runs into extreme difficulties
for the following reasons: (i) if the nonlinear parameters are varied,
it is difficult to optimize them, (ii) if they are not, then the
number of terms required may be excessively large, and, in both cases,
(iii) the properly symmetrized trial function becomes extremely involved.
We show, however, that, instead of performing an optimization, it is
expedient to choose these parameter sets randomly and keep or
discard them by trial and error.
The original procedure of the
SVM, proposed in \cite{SVM}, has recently been greatly developed
with much success for
a fully microscopic multicluster description of light exotic nuclei,
such as $^6$He=$\alpha+n+n$,
$^8$He=$\alpha+n+n+n+n$, $^9$Li=$\alpha+t+n+n$, and
$^9$C=$\alpha+^3$He+$p+p$ \cite{VSL,SVAO}.
\par\indent
Conventional methods \cite{crcbgv,temper} for the choice of the Gaussian
parameters lead to prohibitively large bases for more than 3 or 4 particles.
However, due to the non-orthogonality of the basis
functions, there are different sets of $\underline{a}$ that represent the
wave function equally well. This property of the trial function enables one to
select the most appropriate parameters randomly.
As the inclusion of a new basis state always lowers the energy, one may
characterize its ``utility'' by the energy gained by including it in the
basis. We set up the basis step by step by choosing $\underline{a}$ from
a physically important domain of the parameter space. In the first step we
select a number of parameter sets ${\underline a}$ randomly, and keep
the one that gives the lowest expectation value for the energy.
Next we generate a new random parameter set and calculate the energy
gained by including it in the basis, together with the first basis state. If
the energy gain is larger than a preset value, $\epsilon$, then we admit this
state to the basis, otherwise we discard it and try a new random candidate.
The basis is built up by repeating this until the energy converges.
The rate of convergence can be controlled by dynamically decreasing the
value of $\epsilon$ during the search.
This procedure, although not a full  optimization,
results in very good and relatively small bases.
A similar procedure, called ''stochastic diagonalization'' has been used to
determine the smallest eigenvalue of extremely large matrices \cite{SD}.
\par\indent
The computational cost of this method is moderate.
Having a diagonalization in the $K$th step, the inclusion of the $(K+1)$th
element results in a Hamiltonian matrix whose elements are only nonzero
in the $(K+1)$th row and column and in the diagonal. The lowest eigenvalue
of this simple matrix, the only one required for judging the utility
of a new candidate, can be calculated by a trivial explicit formula.
When a suitable $(K+1)$th basis state has been found, the Hamiltonian
matrix is to be rediagonalized, but that is also relatively simple
if we take advantage of its special form. Thus the random selection process
does not involve a great number of time consuming diagonalizations. Most
computing time is spent on the evaluation of the matrix elements.
\par\indent
We have established a unified and systematic method of
evaluating matrix elements of the trial function. The calculation of matrix
elements is fully analytical and is achieved in three steps: (i) evaluation
of the matrix elements of the few-body system in a single particle (s.p.)
generator-coordinate representation, (ii) transformation from the
s.p. generator
coordinates to Jacobi generator coordinates, and (iii) integral transformation
from the generator coordinates to the parameters $\underline{a}$ of the
correlated Gaussian basis. The s.p. wave functions used in step (i)
are
Gaussian wave packets specified by the generator coordinate {\bf s}
\begin{equation}
\phi ({\bf r},{\bf s})=\left( {2\nu}/{\pi} \right)^{\frac{3}{4}}
\exp\left[-\nu ({\bf r} - {\bf s})^{2}\right] \chi_{\sigma \tau},
\end{equation}
where $\chi_{\sigma \tau}$ is a spin-isospin function.
An $N$-particle
Slater determinant (or, for bosons, a ``Slater permanent")
built up from the s.p. states of eq.~(2) serves as generating function
for the basis function of eq. (1).
The matrix elements of such Slater determinants can be given in closed
analytic forms, which enables us to accomplish the next two steps.
The second step is an orthogonal transformation on the set of generator
coordinates. With this transformation, the dependence on the
center-of-mass generator coordinate factors out, and by omitting this factor,
the center-of-mass motion itself is eliminated. The integral tranformation
between the Jacobi vectors $\{{\bf s}_1,...,{\bf s}_{N-1}\}$ and
$\underline{a}$ is similar to that given by \cite{Kami}.
For the potential-energy matrix elements, first a generic form is evaluated
by assuming the spatial part of the two-particle interaction to be
$\delta (|{\bf r}_j-{\bf r}_i|-r)$.
The matrix element after  the three-step procedure then consists of terms
of the form of
$D(r)\sim r^k {\rm e}^{-p r^2}$, and from these the matrix element
of any $V(r)$ is obtained by performing
$\int_{0}^{\infty} D(r) V(r) dr$. Depending on the actual formula of
$V(r)$, this last integral can either be performed analytically or just
numerically, but the latter is equally fast and accurate.
The dependence of the matrix elements on the nonlinear
variational parameters being known, one can organize the numerical
calculations involved in the random search economically. A change
of the values of the nonlinear parameters does not require
a recalculation of the whole matrix element. Once they have been
calculated for one set of values, to calculate them for many more requires
virtually no time.
The details of the evaluation of the matrix elements will be published
elsewhere.
\par\indent
To assure positive definiteness, ${\underline a}$ is expressed as
${\underline a}={\underline u}$\hspace{2pt}${\underline d}$\hspace{2pt}$
{\underline u}^{t}$,
where ${\underline u}$ is an $(N-1)\times (N-1)$ orthogonal
matrix containing $(N-1)(N-2)/2$ parameters and
${\underline d}$ is a diagonal matrix with $N-1$ positive parameters.
Although no restriction on the
parameters of the orthogonal matrix ${\underline u}$ is necessary,
we found that those connecting different sets of Jacobi-coordinate
systems are especially suitable, and used these matrices in the calculation.
These special ``rotations" describe spatial correlations between
the particles.
In the following we show tables for the ground-state
energies $E$ and point-matter
root-mean-square (rms) radii $\langle r^2\rangle^{1/2}$ calculated with the SVM
for some few-body systems with realistic or model interactions. We shall
compare them with experimental data or with other numerical results for the
same model. The basis dimensions $K$ of the SVM listed in the tables
are those beyond which the energies and the radii do not change in the digits
shown. Each calculation was repeated several times to confirm the convergence.
The average computational time is 10 minutes for a four-body  and
2 hours for a six-body calculation on the VPP500 computer of RIKEN.

The spin-averaged Malfliet--Tjon potential \cite{mt} is most often used for
testing few-body methods. It is a sum of two Yukawa potentials. In Table I we
compare the energies calculated by the SVM for some
$N$=2--6-nucleon systems with those obtained by other methods. The nice
agreement for $N$=3 and 4 corroborates that the SVM is as accurate as
the direct solution of the Faddeev equations \cite{gloeckle,faddeev}
or the method of
the Amalgamation of Two-body correlations into Multiple Scattering (ATMS)
\cite{atms}. The basis used in the Coupled Rearrangement Channel Gaussian
Basis Variational
(CRCGBV) method \cite{crcbgv} is similar to that of the SVM but
the Gaussian parameters are chosen to follow geometric progressions.
The fact that the basis size needed in the SVM is much smaller
proves the efficiency of basis selection in the SVM.
The Malfliet--Tjon  potential, correctly, renders the five-nucleon system
unbound, but, since it is non-saturating, it strongly overbinds $^6$He and,
accordingly, compresses it.
\par\indent
In Table II we show results for the Minnesota potential \cite{minesota}
which is a central interaction, of Gaussian form, containing space-, spin-, and
isospin-exchange operators. The Coulomb interaction between protons is also
included. This potential has often been used in cluster-model
calculations of light nuclei.
Since the method has proved to be accurate and reliable
for the Malfliet--Tjon potential, it is justifiable to view
these calculations as testing the interaction rather than the method,
and therefore, the results are compared with experimental data.
All possible spin and isospin configurations
belonging to the total spin and isospin quantum numbers $S$ and $T$
are allowed for in the trial function and all spherical harmonics that give
nonnegligible contribution are included in $\theta_{LM_L}({\bf x})$.
The energy and size obtained for the triton and for the $\alpha$-particle
converge to values that are close to the experimental data with rather small
bases. The
Minnesota potential does not bind the $N$=5 systems, but it binds $^6$He and
slightly overbinds $^6$Li. $^6$He is found to be much larger than
$^4$He, showing the halo structure of $^6$He. It is important
to note that, for the first time in the application of the Minnesota force,
these results are obtained without assuming any cluster structure or
restricting the model space by any other bias. The agreement is surprisingly
good not only with experiment but also with cluster-model calculations for all
nuclei in which the $N$=2--4 systems are described in terms of a few
0s Slater determinants of different sizes \cite{LKBD}.
The nuclear structure aspects emerging from the present calculations, such as
$\alpha$-clustering in $^6$Li and
the neutron halo in $^6$He, will be discussed in detail elsewhere.
The evaluation of the matrix elements of {\it realistic}
nucleon-nucleon potentials poses no serious problem, and
such calculations are under way.
\par\indent
As is shown in Table III, the method works for pure Coulomb interaction as
well. For the ground state of
the negative positronium ion our calculation reproduces the first six digits
of the extremely accurate variational calculation of Ref.~\cite{epe}, and the
rms radius also agrees with that of the former calculation. For
the dipositronium molecule $(2e^+,2e^-)$ our result is slightly better
than the energy calculated variationally in Ref.~\cite{epep} by optimizing the
same type of basis with the same dimension. This obviously shows
the impracticability of a full optimization of so many non-linear parameters,
and reinforces that the random selection of the parameters is more
powerful. We found no bound states for the ($3e^+,2e^-$) and ($3e^+,3e^-$)
systems. The energy of ($3e^+,3e^-$), for example,
converges to the sum of the energy of a dipositronium molecule and of
a positronium (0.515989 a.u.+0.25 a.u.=0.765989 a.u.), with the
rms radius tending to very large values. The system of
a negative and a positive positronium ion thus forms no bound state but
dissociates into a dipositronium molecule and a positronium.
This result entails that the Coulomb force
cannot bind more than four out of identical charged fermions
and their antiparticles.
\par\indent
To examine the role of the Pauli principle in preventing five negative and
positive electrons from forming bound states, we repeated the same calculation
replacing the fermions by bosonic equivalents.
On a different scale, these systems may be identified, e.g., by
systems of $\pi^-$ and $\pi^+$ with their strong interaction neglected
\cite{boson}. Such bosons turn out to form bound states even in the case of
five particles. As may have been expected, the radius of the charged boson
system decreases by increasing the number of particles.
\par\indent
We also show results for bosonic and fermionic systems with a purely
attractive $Gm^2/r$ (``gravitational") interaction. Self-gravitating
boson systems have
recently attracted some interest \cite{selfgrav}. For these systems,
both variational lower and upper bounds are now available.
In this case even the five-fermion system
is bound. This example shows that the lack of bound states in the
five-electron-positron systems is due partly to the antisymmetry,
and partly to the repulsion between identical particles. As the force
is attractive, the binding energy of the boson systems rapidly increases
with the number of particles ($\sim N(N-1)$).
\par\indent
The limitations of the present method are those implied by the basis size.
The limitations may become excessive as the number of particles
and/or spin and isospin configurations become large.
\par\indent
In summary, we have presented a powerful method for solving bound-state
few-body problems.
Several systems and interactions have been tested, and
accurate results have been obtained. The stochastic variational method
has proved an efficient and economical procedure to find a suitable
variational basis. The unified framework for the evaluation of matrix elements
reported here enables one to treat a great variety of systems and
interactions.

\bigskip

This work was supported by OTKA grants No. 3010 and No. F4348 (Hungary) and by
a Grant-in Aid for Scientific Research (No. 05243102 and No. 06640381) of the
Ministry of Education, Science and Culture (Japan). K. V. gratefully
acknowledges the hospitality of the RIKEN LINAC Laboratory and the
support of the Science and Technology Agency of Japan. We thank
Prof. R. G. Lovas for careful reading of the manuscript.

\begin{table}
\caption{Energies and point-matter rms radii
of $N$=2--6-nucleon systems interacting via the Malfliet-Tjon
potential V [14].}

\begin{tabular}{llddr}
$N$ & Method                  & $E$ (MeV) & $\langle r^2\rangle^{1/2}$ (fm) &
$K$\\
\hline
2     & Numerical               & $-$0.4107    &3.743    & \\
         & SVM                   & $-$0.4107    &3.743    & 5  \\
\hline
3     & Faddeev\tablenote{Ref. 4.}            & $-$8.2527   &   &    \\
         & SVM                                     & $-$8.2527 & 1.682  & 80 \\
\hline
4     & ATMS\tablenote{Ref. 2.}                & $-$31.36  &            \\
       & CRCGBV\tablenote{Ref. 1.}       & $-$31.357 &       & 1000 \\
       & FY\tablenote{Ref. 3.}       & $-$31.36 &       &     \\
       & SVM                         & $-$31.360 & 1.4087  &  150 \\
\hline
5     & SVM                         & unbound &      &  \\
\hline
6 ($^6$He) & SVM                         & $-$66.30 & 1.52   & 800   \\
\end{tabular}
\end{table}

\begin{table}
\caption{Energies and point-matter rms radii of few-nucleon systems with
the Minnesota potential [15] with exchange parameter $u=1$. The Coulomb
interaction is included. The experimental $\langle r^2\rangle^{1/2}$
value is the charge radius with the proton size corrected.}

\begin{tabular}{llddr}
$N$ & Method                  & $E$ (MeV) & $\langle r^2\rangle^{1/2}$
(fm) & $K$\\
\hline
2     & SVM                   & $-$2.202    &1.952    & 5  \\
       & Exp.            & $-$2.224    &1.96    &   \\
\hline
3 ($^3$H)    & SVM                     & $-$8.380   & 1.698  & 40 \\
       & Exp.                           & $-$8.481    & 1.57  &   \\
\hline
4 ($^4$He) & SVM              & $-$29.937 & 1.41     &  60 \\
            & Exp.       & $-$28.295       & 1.47        &     \\
\hline
5     & SVM                          & unbound &      &  \\
       & Exp.                         & unbound &      &  \\
\hline
6 ($^6$He) & SVM                         & $-$30.07 & 2.44  & 600   \\
            & Exp.                  & $-$29.271 &    &     \\
\ \   ($^6$Li) & SVM                         & $-$34.59 & 2.22  & 600   \\
            & Exp.                  & $-$31.995 & 2.43   &     \\
\end{tabular}
\end{table}

\begin{table}
\caption{Energies and point-matter rms
radii of charged electron-positron systems treated as fermions (f) and
as bosons (b). Atomic units are used.}
\begin{tabular}{llddr}
System & Method                  & $E$  & $\langle r^2\rangle^{1/2}$ & $K$\\
\hline
($e^+,e^-$) b,f   & SVM                   & $-$0.25  &1.732    & 10  \\
          & exact        & $-$0.25    &1.732    &     \\
\hline
($2e^+,e^-$) b,f   & SVM                  & $-$0.262004   & 4.592       & 150
\\
            & Var.\tablenote{Ref. 17.}   & $-$0.26200507  & 4.594  & 700  \\
\hline
($2e^+,2e^-$) b,f  & SVM           &$-$0.515989  & 3.608 &  300 \\
            & Var.\tablenote{Ref. 18.}     & $-$0.515980  & 3.600 &  300 \\
\hline
($3e^+,2e^-$) f   & SVM                         & unbound  &  & 1000  \\
($3e^+,2e^-$) b   & SVM                         & $-$0.5493 & 3.53  & 200  \\
\hline
($3e^+,3e^-$)  f & SVM                         & unbound &  & 1000     \\
($3e^+,3e^-$)  b & SVM                         & $-$0.820 & 3.42 & 300   \\
                & Var.\tablenote{Ref. 19.}       & $-$0.789 &      &   5   \\
\end{tabular}
\end{table}

\begin{table}
\caption{Energies and point-matter rms
radii of ``self-gravitating" $m$-particle--$n$-antiparticle systems
($m+,n-$); f: fermions; b: bosons.
VLB and VUB stand for the variational lower
and upper bounds given in Ref. [20]. The units of the energy and length
are $G^2 m^5\hbar^{-2}$ and  $G^{-1} m^{-3} \hbar^2$, respectively.}
\begin{tabular}{llddlr}
System           & Method      & $E$      &           &
\hspace{-5pt}$\langle r^2\rangle^{1/2}$
  & $K$\\
\hline
($+,-$) b,f  & SVM         & $-$0.25  &\hspace{-8pt}
&\hspace{-5pt}1.732    & 10  \\
                 & exact       & $-$0.25  &\hspace{-8pt}
 &\hspace{-5pt}1.732    &     \\
\hline
($2+,-$) b,f & SVM         & $-$1.072 &\hspace{-8pt}
&\hspace{-5pt}1.304& 15 \\
                 &
       Var.\tablenote{Ref. 20} & $-$1.067 &\hspace{-8pt}           &      &
\\
\hline
($2+,2-$) b,f& SVM         & $-$2.791 &\hspace{-8pt}
&\hspace{-5pt}1.027&  100 \\
                 & VUB \ (VLB) & $-$1.951 &\hspace{-8pt} ($-$3.00) &      &  \\
\hline
($3+,2-$) f  & SVM         & $-$3.758 &\hspace{-8pt}
&\hspace{-5pt}1.554& 200  \\
($3+,2-$) b  & SVM         & $-$5.732 &\hspace{-8pt}
&\hspace{-5pt}0.844& 200  \\
                 & VUB \ (VLB) & $-$4.336 &\hspace{-8pt} ($-$6.25) &      &  \\
\hline
($3+,3-$) f  & SVM         & $-$6.409 &\hspace{-8pt}
&\hspace{-5pt}1.621& 300   \\
($3+,3-$) b  & SVM         & $-$10.215&\hspace{-8pt}
&\hspace{-5pt}0.718& 300 \\
                 & VUB \ (VLB) & $-$8.130 &\hspace{-8pt} ($-$11.25)&      &
 \\
\end{tabular}
\end{table}

\end{document}